\documentclass[a4paper]{revtex4}
\usepackage{graphicx}
\usepackage{fancyhdr}
\usepackage{amsmath}
\usepackage{color}

\begin{document}

\title{$Z'$ search in non-minimal Universal Extra Dimensions: two bumps and interference}

%

\author{{AseshKrishna} Datta${}^{\dagger}$, Kenji Nishiwaki${}^{\ddagger}$\footnote{This poster presentation was given by Kenji Nishiwaki in the conference HPNP2015.}, Saurabh Niyogi${}^{\P}$}
\affiliation{
${}^{\dagger}${Harish}-Chandra Research Institute, Allahabad 211-019, India\\
${}^{\ddagger}$School of Physics, Korea Institute for Advanced Study, Seoul 130-722, Republic of Korea\\
${}^{\P}$Institute of Mathematical Sciences, Chennai 600-113, India
}

%
%

\begin{abstract}
We discuss prospects of the $Z'$ search at the LHC in non-minimal Universal Extra Dimensions with tree-level brane-local terms in five dimensions. In this scenario, we {find two} major differences from {the usual} $Z'$ physics: (i) two $Z'$ candidates {close-by in mass exist}; (ii) the effective couplings to the SM fermions could be very large due to drastic overlapping of their profiles along the extra dimension. To evaluate the actual situation precisely, we reconsider the important issues of resonant processes, {\it i.e.}, treatment of resonant propagators and including interference effects.
(This talk is based on tentative results of an ongoing project.)
\end{abstract}

\maketitle

\thispagestyle{fancy}


\section{Introduction}
{Appearance of new heavy vector bosons is common to many physics scenarios beyond the Standard Model~(SM).}
{Thus, search for these states is a standard program in the hunt for new physics at colliders.}
Especially, signals of the resonant production of {electromagnetically} neutral ones ($Z'$ candidates) {leading to a pair of leptons} are very clean and {hence very} useful.
These neutral bosons are associated with a new gauge group which is spontaneously broken at {the} scale of new physics, {\it e.g.}, {the} Grand Unified theories, {the} left-right symmetric models or {the} $U(1)$ extensions of the SM around the electroweak scale.
Another origin of new vector bosons is (compactified) extra dimensions.
When {a higher-dimensional gauge boson lives} in the bulk space, various massive copies of the lightest mode knows as Kaluza-Klein~(KK) particles appear in {four-dimensional compactified} theories.
{Well-known} possibilities {in this class of theories with} five-dimensional~(5D) spacetime are the bulk SM in the Randall-Sundrum warped background or the Universal Extra Dimension~(UED) scenarios~\cite{Appelquist:2000nn} (mainly) in the flat space.
In such cases, 5D bulk {$SU(2)_W$} and $U(1)_Y$ gauge bosons are found, where {there are} two types of $Z'$ candidates.
Note that when an accidental $Z_2$ parity originating from a reflection in the extra space {ensures the existence of a dark matter candidate in the system}, level-1 massive bosons cannot decay into a pair of SM particles, {while the level-2 gauge bosons could do.}
Therefore, typically level-2 gauge boson {resonances in UED are sensitive to such Drell-Yan type search.}

In the minimal UED case without tree-level brane-local terms, interactions between level-2 gauge bosons and the SM fermions are generated at the one-loop level~\cite{Cheng:2002iz}.
{Hence}, two sharp nearby resonances are predicted, which can be {the `smoking gun'} signals for the UED {type scenarios}~\cite{Datta:2005zs}.
On the other hand, the expected production cross section is not large since this process is {a} loop-induced one.
Also, to evade the recent bound on {level-1} colored particles at $7$ and $8$ TeV {runs of the} Large Hadron Collider~(LHC),
the typical scale of (level-1) KK particles, which is the inverse of the radius of the extra dimension, should be roughly {larger} than $1 \sim 1.5$ TeV.
Thereby, it is not so easy to probe the minimal UED scenario through this channel {via level-2 states of such a kind}.

When we introduce (tree-level) brane-local terms in the minimal scenario, {situations} can be drastically changed~\cite{Carena:2002me,delAguila:2003bh,Flacke:2008ne,Datta:2013yaa}.
Thanks to brane-local kinetic terms~(BLKTs), we can {have situations} where interactions between {the} level-2 gauge bosons and the SM fermions are present at the tree level and relatively light $Z'$ candidates {$(B_{(2)}, W^3_{(2)})$} are possible.
Interestingly, at least a part of the parameter space {in} which the above properties are realized, the effective gauge interaction is enhanced via a large overlapping of wave functions of the particles along the extra direction.
{Thus}, $Z'$ resonant production {becomes a viable} channel for probing {such a} non-minimal UED~{(nmUED)} scenario at the LHC.

Here, a nontrivial feature is expected.
When effective gauge couplings between a $Z'$ candidate and the SM fermions take large values, the $Z'$ resonance becomes broad.
{In such a situation, a fresh look into the validity of the narrow-width approximation, which is commonly used in studies of (narrow) resonances, is needed.}

\section{Properties of $Z'$ candidates in non-minimal UED models}

\begin{figure}[ht]
\centering
\includegraphics[width=0.35\columnwidth]{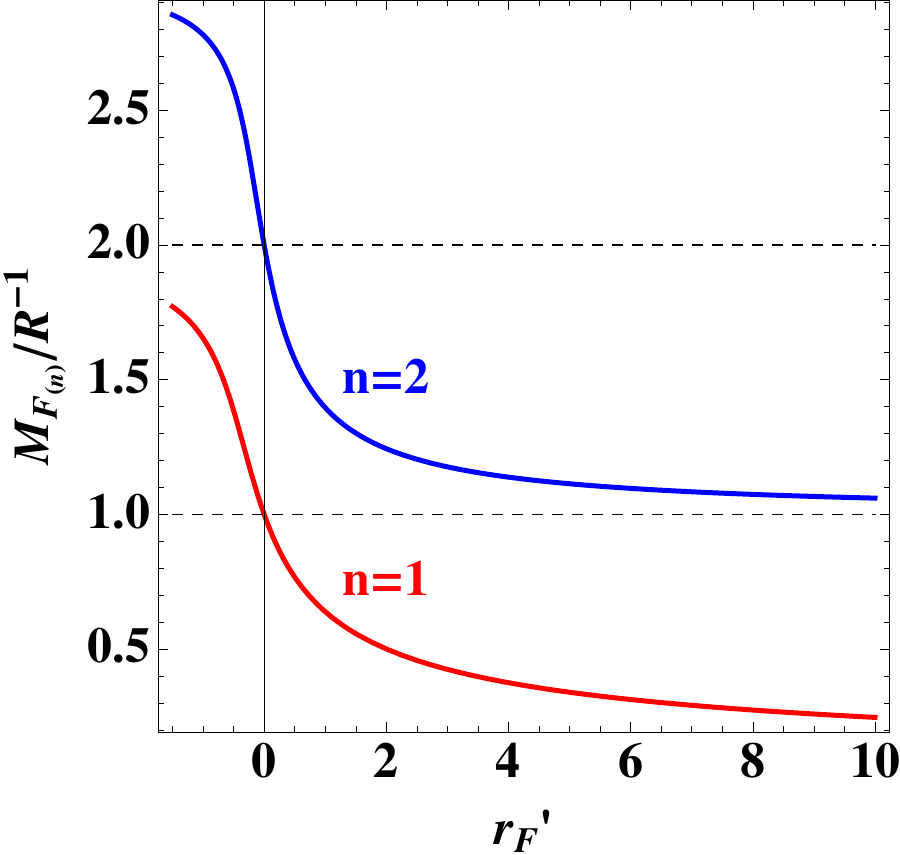}
\qquad
\includegraphics[width=0.35\columnwidth]{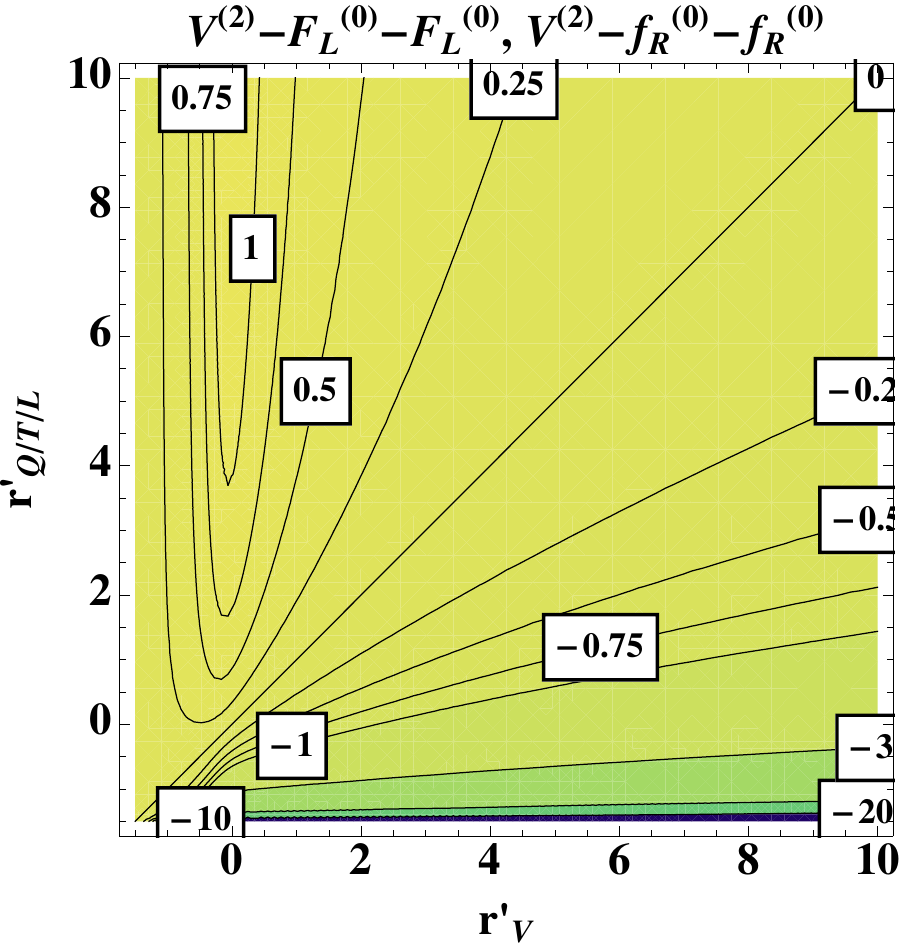}
\caption{
Left: The generic profile of the variation of
$M_{F_{(n)}}/R^{-1}$ as a function of $r'_F \,(= r_F R^{-1})$ for the cases $n=1$ and $n=2$~\cite{Datta:2013yaa}.
Right: Contours of deviation 
for the generic couplings $V^{(2)}$-$F^{(0)}_L$-$F^{(0)}_L$ (or
$V^{(2)}$-$f^{(0)}_R$-$f^{(0)}_R$)
from the 
corresponding SM values in the $r^\prime_V-r^\prime_{Q/T/L}$ plane. 
$V$, $F$ and $f$ stand for generic gauge boson, $SU(2)_W$ doublet and 
singlet fermion fields (with corresponding chiralities), respectively.
Note that when $V$ is the (KK) $W$ boson, types of the two 
fermions involved at a given vertex are different~\cite{Datta:2013yaa}.
}
\label{plot1}
\end{figure}

In the non-minimal UED scenarios with BLKTs, the effective KK mass of a level-$n$ KK state $F_{(n)}$ {is different} from the {corresponding} value in the minimal case (at the tree level), $M_{F_{(n)}} = n/R\,(n=1,2,3,\cdots)$, {where $R$ is the radius of the extra dimension.}
$r'_F (\equiv r_F R^{-1})$ is the {dimensionless} coefficient {corresponding to} the brane-local kinetic term $r_F$ for 5D field $F$.
{It plays an important role in phenomenology of the non-minimal scenarios.}
As shown in the left panel of Fig.~\ref{plot1},
{KK masses different from these in the minimal scenario can be obtained with varying $r'_F$, where each 5D field can take an individual coefficient leading to wide varieties of mass spectra.}

In general, $r_F$ also affects the strengths of {the} effective gauge interactions {involving} KK particles $F_{(n)}$.
In particular, the case of $V^{(2)}$-$F^{(0)}_L$-$F^{(0)}_L$ (or $V^{(2)}$-$f^{(0)}_R$-$f^{(0)}_R$) is interesting, where $V$, $F$ and $f$ stand for generic gauge boson, $SU(2)_W$ doublet and 
singlet fermion fields (with corresponding chiralities), respectively.
The right panel of Fig.~\ref{plot1} indicates the region {where} significant coupling enhancement is found when the {dimensionless coefficients of} BLKTs for 5D fermions {($r'_{Q/T/L}$)} are negative.
{With suitable} coefficients for {the} electroweak gauge bosons, two relatively lighter $Z'$ candidates are realized with large effective couplings to the SM fermions.
Therefore, the $Z'$ search at the LHC {can probe a favorable region of the nmUED parameter space}.

\section{Propagator schemes}

Usually in studies {of} resonant production of a particle with a narrow width,
{one adopts} the narrow-width approximation.
{This} means that we use the simple form of a resonant propagator generating the Breit-Wigner~(BW) shape in the resonant mass distribution
\begin{align}
\frac{1}{(\hat{s} - m_R^2) + i m_R \Gamma_R},
	\label{BW_prop}
\end{align}
where $\hat{s}$ indicates the total energy at the parton level, $m_R$ and $\Gamma_R$ represent the mass and the total width of the {resonant state}, respectively.
A more precise form adopted in {a widely used event generator like} {{Pythia\,6}~\cite{Sjostrand:2006za}} is as follows:
\begin{align}
\frac{1}{(\hat{s} - m_R^2) + i \sqrt{\hat{s}} \Gamma_R(\hat{s})},
\end{align}
where the total width is replaced to energy-dependent one and the imaginary part of the two-point function of the resonant particle is correctly described.
The full form of the propagator is
\begin{align}
\frac{1}
{\left(1 - \mathcal{R}\left[\Pi'_T(m^2)\right]\right) \left(\hat{s} - m^2\right) + \Pi_T(\hat{s}) - \mathcal{R}\left[\Pi_T(m^2)\right]},
	\label{full_prop}
\end{align}
where $\mathcal{R}$ {stands for the} real part, $\Pi_T(\hat{s})$ is the renormalized {(transverse part)} two-point function (of the gauge boson) in the on-shell scheme and $m$ stands for the (renormalized) pole mass.
In this {way}, not only the imaginary part but also the real part are fully included.

Generally for large widths $(\Gamma/M > 10 \sim 15\%)$, it is nontrivial {to estimate the resulting effects and hence to work out at the colliders.}
Our interest is just in such {scenarios} and we carefully treat this issue in our analysis.

\section{Tentative results}

In this {section}, we {present} our {\it tentative} results on future LHC reach at $\sqrt{s} = 13\,\text{TeV}$ {for} the non-minimal UED scenarios with/without $Z'$ coupling enhancement. 
{We work with} four benchmark points of the universal BLKTs for 5D quarks and leptons $r'_{Q} = r'_{L} = -1.0,\, -0.5,\, 0.0,\, +0.5$ with $R^{-1} = 3\,\text{TeV}$ and $r'_{SU(2)_L} = r'_{U(1)_Y} = 1.3$.
Here, we assume {two things}: (i) the Weinberg angle {between} the two states are exactly zero, which would be realized after considering one-loop corrections~\cite{Cheng:2002iz}; (ii) there is a $150\,\text{GeV}$ {mass-split} between the two massive gauge bosons $B_{(2)}$ and $W^{3}_{(2)}$ like in the corresponding minimal UED case, consequently the masses of the two level-2 states are
\begin{align}
m_{B_{(2)}} \simeq 4.0\,\text{TeV},\quad  m_{W^{3}_{(2)}} \simeq 4.15\,\text{TeV}.
\end{align}
For collider simulations, we use Pythia\,6.4~\cite{Sjostrand:2006za} with our modifications for implementing the BW and the ``full" propagators shown in Eqs.~(\ref{BW_prop}) and (\ref{full_prop}), respectively.
CTEQ6L parton distribution function~\cite{Pumplin:2002vw} is adopted and the factorization and renormalization scales are set at $m_{B_{(2)}}$.
We only focus on the signal process with dimuon final state, $p p \to B_{(2)}/W^{3}_{(2)} \to \mu^+\mu^-$, while the major background process within the SM, $p p \to \gamma^\ast/Z \to \mu^+ \mu^-$ is considered.
The following kinematic cuts are imposed:
\begin{itemize}
\item $p_{T}^{\mu} > 30\,\text{GeV}$,
\item $\eta^{\mu} < 2.5$,
\item $M_{\mu^+ \mu^-} > 2\,\text{TeV}$: {the} flat cut on the invariant mass on the muon pair. This choice may not be optimal, {and can be improved for specific masses of the resonances being probed.}
\end{itemize}
Note that we ignore detector effects for simplicity.
To estimate the reach, we adopt a rather strict set of criteria: we require a minimum of five events with a significance $S \geq 5\sigma$.
We use the following Poisson significance estimator formula in Ref.~\cite{Cowan:2010js}
\begin{align}
S = \sqrt{ 2 \left[ \left( s+b \right) \log\left( 1 + \frac{s}{b} \right) -s   \right] },
\end{align}
where $b$ and $s$ are the expected numbers of background and (background-subtracted) signal events after the cuts, respectively.

Results are summarized in table~\ref{result_table}, where ``scale factors" represent effective $Z'$ coupling deviations from the SM $Z$ {(sequential $Z'$)} ones defined as ratios and the total widths of $B_{(2)}$ and $W^{3}_{(2)}$ are calculated by Madgraph\,5~\cite{Alwall:2011uj} with our own model file generated via FeynRules~\cite{Alloul:2013bka}.
Note that in the ``full" propagator scheme, we consider the complete treatment only for the SM fermions.
For other KK particles, we simply adopt the BW treatment.
Also, we analyze the process with and without interference effects between two nearby resonances.
Here, in the ``large" coupling cases ($r'_{Q} = r'_{L} = -1.0,\, -0.5$), we can see that propagator treatments and interference {affect} {the} required luminosities for discovery {($10 - 20\%$)}.
{It is equally relevant to express this effect in terms of the ``degraded" reach in $m_{Z'}$ for a given luminosity.}

\begin{table}[ht]
\begin{center}
\caption{{Integrated luminosities (in fb{}$^{-1}$)} required for the discovery of level-2 $Z'$ candidates evaluated both in the cases of the BW propagator and propagators with full one-loop corrections of the SM fermions.
``w int" and ``wo int" stand for ``with interference" and ``without interference" of the two resonant {states}, respectively.}
\begin{tabular}{|c||c|c|c|c||c|c|} \hline
$r'_{Q},\, r'_{L}$ (scale factor) & full, w int & BW w int & full wo int & BW wo int & $B_{(2)}$ Width (GeV) & $W_{(2)}^3$ Width (GeV) \\ \hline
$-1.0$ ($-3.5$) & $0.352$ & $0.411$ & $0.368$ & $0.43$ & $826$ & $1805$ \\ \hline
$-0.5$ ($-1.46$) & $5.32$ & $5.46$ & $6.22$ & $6.38$ & $154$ & $397$ \\ \hline
$0.0$ ($-0.72$) & $108$ & $109.5$ & $109.7$ & $115.7$ & $47$ & $173$ \\ \hline
$0.5$ ($-0.34$) & $8516$ & $8601$ & $8611$ & $8765$ & $20$ & $116$ \\
\hline
\end{tabular}
\label{result_table}
\end{center}
\end{table}

\section{Summary}

We have discussed {the $Z'$-like} physics of the non-minimal UED model with tree-level brane-local terms, where the masses of the two $Z'$ candidates can get to be relatively light compared {to} the typical value {of $2 R^{-1}$} in the minimal case, and large {enhancements could take place in the effective couplings to the SM fermions} in a part of the parameter space.
Another significant feature is that the two resonant peaks are close to each other.
In such a situation, correct treatments of {the} resonant propagators and interference effects among the two {nearby states could generate} sizable effects {($10 \sim 20\%$ deviations)} in required luminosities for {the} discovery {at the LHC}.

\section*{Acknowledgement}

{KN would like to thank Shinya Kanemura and all the other members of the organizing committee of HPNP2015
for the stimulating opportunity and their kind hospitality.}
KN and SN were partially supported by funding available from the 
Department of Atomic Energy, Government of India for the Regional 
Centre for Accelerator-based Particle Physics (RECAPP), Harish-Chandra 
Research Institute.
KN would like to thank Genevi\`eve B\'elanger, Giacomo Cacciapaglia, Aldo Deandrea and Tetiana Berger-Hryn'ova for useful discussions.
KN also thank Laboratoire d'Annecy-le-Vieux de Physique Th\'eorique and Universit\'e de Lyon for {the} kind hospitality.


\bigskip 

\end{document}